\documentstyle[aps,preprint,psfig]{revtex}
\begin{document}

\title{Circular Kinks on the Surface of Granular Material Rotated in a
Tilted Spinning Bucket} 

\author{Sangsoo Yoon$^{1,2}$, Byeong-ho Eom$^{1,2}$, Jysoo Lee$^1$ and
$^{\dagger}$Insuk Yu$^{1,2}$} \address{Department of Physics$^1$ and
Condensed Matter Research Institute$^2$\\ Seoul National University,
Seoul 151-742, Korea} 
\maketitle 

\begin{abstract}
We find that circular kinks form on the surface of granular material
when the axis of rotation is tilted more than the angle of internal
friction of the material. Radius of the kinks is measured as a
function of the spinning speed and the tilting angle. Stability
consideration of the surface results in an explanation that the kink
is a boundary between the inner unstable and outer stable regions. A
simple cellular automata model also displays kinks at the stability
boundary.
\end{abstract}
\pacs[PACS number: 46.10.+z, 47.20.Ma, 47.54.+r, 81.05.Rm]

Granular materials behave differently from any other familiar forms of
matter.  They possess both solid- and fluid-like nature and exhibit
unusual dynamic behaviors, such as segregation, surface waves, heap
formation and
convection\cite{Jaeger,Hayagawa,Mehta1,Powders,Campbell}. The surface
of granular material in a spinning bucket is an example of such
interesting phenomena. Vavrek and Baxter\cite{Vavrek} showed that the
surface shape of sand in a vertical spinning bucket can largely be
explained using Coulomb's criterion. Medina {\it et al}.\cite{Medina}
investigated hysteresis of the surface shape with semi two-dimensional
rotating bins and showed the existence of multiple steady-state
solutions. Yeung\cite{Yeung} studied the system with an initially
conical surface. By using a model of granular surface
flow\cite{Bouchaud,Mehta2}, he found that the behaviors of the model
agreed well with the experiments.

When we tilt the rotational axis of a bucket from the vertical
direction, circular kinks develop on the granular surface if the
tilting angle is greater than the angle of internal friction of the
material. A glass beaker of 10.5 cm diameter and 1 liter capacity is
used as a spinning bucket. The beaker is mounted on a dc motor to
rotate around its axis of symmetry. The motor is fixed to a stand such
that the tilting angle $\alpha$ can be varied. The rotation rate is
varied from 0 to 300 rpm by controlling the voltage. The angular
velocity $\omega$ is measured by using a photogate timer and is
constant within an error smaller than 1 rpm. We use natural sand, as
used in general construction, as a prototypical granular material. In
order to insure the monodispersity of sands, two sieves with 0.35 and
0.25 mm meshes are used and the sizes in between are selected. The
angle of internal friction (angle of repose) and apparent density are
found to be $\theta_f $ = 34 $\pm$ 1$^{\circ}$ and $\rho$ = 1.52
g/cm$^3$, respectively.

Figure 1(a) is a schematic side-view and 1(b) is a top-view photograph
of a circular kink formed on the sand surface. To measure the diameter
of the kinks, a divider is placed near a kink and matched with the
diameter. Several measurements are carried out for each kink and
errors of measurements are about 1 mm. We also measure the surface
shapes in some cases using the method of Ref.  6.

First, we tilt the bucket by angle $\alpha$ and then turn on the motor
to various speeds. In this type of experiment the initial granular
surfaces are inclined flat surfaces. After a few minutes of rotation,
we measure the radius of kinks $r_k$. The radius depends on $\omega$
as $r_k \propto \omega^{-2+x} $ (\(-0.2 \leq x \leq 0.1\)), thus a
dimensionless radius $R_k = r_k \omega ^2 /g$, which is roughly
constant, can be introduced (Fig. 2). Next, the tilting angle is
varied at fixed $\omega$. When $\alpha \leq 30^{\circ}$, there forms a
paraboloid-like granular surface whose shape depends on the initial
condition but does not change with time. As $\alpha$ becomes larger
than about 35$^{\circ}$, a circular kink forms which is independent of
the initial granular surface. Inside the kink, most of the grains on
the surface avalanche during the rotation and we can see a complex
motion like a whirlpool. The surface shape of the inner region is
asymmetric about the rotation axis. On the other hand, those grains at
and outside the kink are stationary with respect to the bucket. The
surface shape of the solid-like region is in general asymmetric and
depends on the initial surface. However, when we first rotate the
bucket vertically with an initially conical surface and then tilt
slowly, the surface shape remains nearly symmetric. The kink is a
boundary between the two dynamically distinguishable regions. As
$\alpha$ increases, the $R_k $ tends to increase. The measured values
of $R_k $ at several tilting angles are shown in Fig. 3. For $\alpha >
70^{\circ}$, a new type of instability appears; some sand grains are
separated from the surface and fall freely during the rotation.

We also study hysteresis by sequentially increasing, decreasing or
randomly changing the angular velocity. For fixed $\alpha$, the radius
of kinks is determined only by $\omega$ and does not depend on the
past history of changing $\omega$. But the shape of the surface shows
hysteresis. When $\omega$ increases a new kink appears inside the
previous one. The previous kink formed at slower $\omega$ can be
frozen in the solid-like outer region. Thus, there can be many
concentric kinks. On the other hand, when $\omega$ decreases a new
kink appears outside the previous one and the previous kinks are
always washed out.

Various vessels are tested as spinning buckets. If the width of a
container is larger than the diameter of a kink, the radius of the
kink does not depend on the size or shape of the container. We find
that, to form the kinks, it is not necessary for the rotational axis
to coincide with the axis of symmetry of a container. When we rotate a
rectangular box around an axis which is fixed on one of the
side-walls, there appears a semicircular kink.

Similar kinks are also found with the silica-gel (Matrex Silica,
Amicon Corp., Danvers, MA 01923, U.S.A., of apparent density $\rho$ =
0.35 g/cm$^3 $, particle diameter around 0.2 mm, and the angle of
internal friction $\theta_f = 30 \pm 1^{\circ}$), but the $R_k $'s are
smaller by about 20 \% than with the sand. We try sugar powder
(sucrose, grain size $\sim$ 0.2 mm, after 100$^{\circ}$C for 2 hours
to remove moisture), and observe kinks too.

We now discuss the stability of the granular surface. There are four
forces acting on a grain at the surface in the rotating frame: gravity
$m\vec{g}$, centrifugal force $m\vec{r}\omega^2$, normal force
$\vec{N}$ and frictional (shearing) force $\vec{f}$. Here, $m$ is the
mass of a grain, $\vec{g}$ the gravitational acceleration and
$\vec{r}$ is the radial displacement of the grain from the rotational
axis. We make the following assumptions. First, there is no bulk
motion in the pile and the grain at the surface can only slide or roll
on the surface. Second, there is no inertial effect. The grain stops
as soon as it satisfies a force balance equation. Third, we also
assume the Coulomb yield condition \cite{Vavrek,Jackson}, which states
that the grain does not move when $f \leq \mu N$, where $\mu$ $(\mu =
\tan \theta_f )$ is the coefficient of friction. The force balance
equation for the granular system is
\begin{eqnarray}
\vec{N} + m\vec{g} + m\vec{r}\omega^2 +\vec {f} =0,~~f \leq \mu N.
\end{eqnarray}

We define a dimensionless vector
\begin{eqnarray}
\vec{n}_h ~=~- \frac{m\vec{g}+m\vec{r}\omega^2}{m g},
\end{eqnarray}
which is normal to the stable surface when there is no friction
($\vec{f}~=~0$). The inclination of the rotational axis breaks the
cylindrical symmetry of the vector $\vec{n}_h$ and makes the angle
$\beta$ between $\vec{n}_h $ and $\hat{z}$ time-dependent, where
$\beta$ is also the angle between the stable surface without friction
and the bottom plane of the spinning bucket (Fig. 4).

The kink is an abrupt change in the slope of the granular surface
along the radial direction. Therefore, we concentrate on the radial
motion of grains. The radial stability condition can be expressed in
terms of the angle $\beta$
\begin{eqnarray}
\tan( \beta(t) - \theta_f )~\leq~ \tan \theta ~ \leq ~\tan( \beta(t) +
\theta_f ),
\end{eqnarray}
where $\tan \theta$ is the local slope of the granular surface in the
radial direction with respect to the bottom of the bucket (Fig.  4).

The stable surface satisfying Eq. (3) for all $t$, can exist only if
the following condition is satisfied;
\begin{eqnarray}
\beta_{\mbox{max}} - \theta_f ~\leq~ \beta_{\mbox{min}} + \theta_f ,
\end{eqnarray}
where $\beta_{\mbox{max}}$ ($\beta_{\mbox{min}}$) is the maximum
(minimum) value of $\beta(t)$ during rotation. If we use a cylindrical
coordinate system ($\rho, \phi, z$) aligned to the rotation axis, the
$\vec{n}_h $ and $\beta (t)$ can be expressed as
\begin{eqnarray}
\vec{n}_h ~=~\hat{\rho}(\sin \alpha \sin
\phi-R)+\hat{\phi}\sin\alpha\cos\phi+\hat{z}\cos \alpha
\end{eqnarray}
and
\begin{eqnarray}
\cos \beta(t)~=~\frac {\cos \alpha }{\sqrt{R^2 + 1 -2R\sin \alpha \sin
\omega t}},
\end{eqnarray}
where $R=r\omega^2 /g$ is a dimensionless radius and $\phi = \omega t$
(the phase is chosen that $\phi = \pi /2$ corresponds to the highest
position during rotation). The $\beta_{\mbox{max}}$
($\beta_{\mbox{min}}$) in Eq. (4) becomes the angle $\beta(t)$ at
$\phi = 3\pi /2$ ($\phi = \pi /2$).

Finally, we reach the following radial stability condition,
\begin{eqnarray}
R~\geq\left\{~\begin{array}{ll} 0 &\mbox{$(\alpha \leq \theta_f)$}\\
 R_c \equiv \sqrt{\frac{\sin2(\alpha- \theta_f )}{\sin2\theta_f }}
 &\mbox{$(\alpha > \theta_f)$}.
\end{array}
\right.
\end{eqnarray}
If $\alpha \leq \theta_f$, the steady state surface is stable for all
$R$. If $\alpha > \theta_f$, the surface can be stable only in the
region where $R$ is larger than the critical radius $R_c$. In the
region with smaller $R$, avalanches of grains occur. We plot the
$R_c(\alpha)$ in Fig. 3 for the sand sample with $\theta_f=34^{\circ}$
to compare with the radius of kink. The critical radii calculated by
Eq. (7) reflect the qualitative features of the experimental kink
radius; they increase with the inclination angle and do not depend on
the angular velocity.

The angular stability condition is also examined. The corresponding
critical radius $R_c ^{\phi}$ for the azimuthal direction is always
larger than $R_c$, and their difference increases monotonically as
$\alpha$ increases. In our experiment ($35^{\circ} \leq \alpha \leq
70^{\circ}$), however, the difference is small and the angular
instability is expected to have little effect on the formation of
kinks.

In order to gain insights on the formation of the kinks, we study the
system using a simple cellular automata model similar to that of Bak
{\it et al}.\cite{Bak1,Bak2}. In addition to the assumptions made
earlier, we further assume that the relative motion of grains in the
azimuthal direction is not significant. Experimentally, grains just
inside the kink show nearly circular trajectories without relative
movement in the azimuthal direction. Since the kinks are formed by the
motion of grains nearby the kinks, we expect that the above assumption
does not alter their formations. The two-dimensional granular surfaces
can now be described by one-dimensional curves--- the surface profile
$h(R)$ at a given $\phi$.

The spatial coordinate $R$ is made discrete---it is replaced by $R_{i}
= i ~\Delta R$ ($i=-n,-n+1,\cdots,n$), which runs from one end of the
container to the other end.  We measure local slopes $s_{i}$, defined
as $(h(i) - h(i+1))/ \Delta R$, and check the stability of all the
slopes following the criterion of Eq. (3). We then update the heights
of the pile to make the local slopes at least marginally stable
starting from the uphill and proceeding towards the downhill.

If the local slope $s_{i-1}$ becomes unstable due to an update at the
$i$-th site, we allow ``backward propagation'', where a perturbation
at a site can influence a site uphill from the perturbation.  To be
more specific, we decrease the height $h(i-1)$ and transfer the excess
amount to the $(i+1)$th site. If the change makes the $(i-2)$th site
unstable, its height is decreased in a similar way. We proceed until
all local slopes become stable. Then the container is rotated by a
small angle $\delta \phi$ and we again update the height. For
computational simplicity we assume that the relaxation time of the
pile is much shorter than the period of the rotation. At the boundary,
we apply a mass conservation condition; grains cannot enter nor leave
the container.

In Fig. 5, we show time evolutions of the heights $h(i,t)$ with
different initial conditions plotted during a half rotation. Here,
$\Delta R=0.01,~\alpha = 45^{\circ},~\theta_f = 34^{\circ}$, and
$\delta \phi = \frac{\pi}{3} \times 10^{-2}$ rad. One can see the
region with large $|R|$ is solid-like; the profile does not change
with time. On the other hand, the inner region is fluid-like; the
heights keep changing. There is a discontinuity in the slope at the
border of the two regions ($|R|\simeq 1.2$). The resulting kink looks
similar to what is observed in the experiments. Although the formation
of the kinks does not depend on the initial surface, the profile of
the solid-like region does.  The radii $R_k(\alpha)$ resulting from
the simulation with $\theta_f = 34^{\circ}$ are shown in Fig.  3 for
different $\alpha$ in the range $(0^{\circ}-90^{\circ})$. One notices
that not only the qualitative features such as the threshold angle
$(\alpha = \theta_f )$ and the overall shape, but also their numerical
values are in good agreement with the measured ones.  However, this
quantitative agreement may be accidental, since we expect the present
simple model to reproduce qualitative, but not quantitative behaviors.
To study the hysteresis behaviors, we start with a kink formed on the
surface at $\alpha= 40^{\circ}$, and suddenly increase $\alpha$ to
$65^{\circ}$. There indeed appears a new kink with smaller radius,
leaving the old one stable, similar to what is observed in the
experiments.

Finally, the formation of the kinks can be viewed as follows. Consider
the border at $R_c$ from the stability analysis. There will be
avalanches just inside the border. Due to the backward propagation,
which we expect to exist in real sandpiles, the region just outside
the border should be involved in the avalanche. This explains why the
measured $R_k$ is larger than $R_c$ as shown in Fig. 3. The surface in
the stable region will be marginally stable, so we expect its slope is
a smooth function of $R$. The surface in the inner unstable region,
however, is produced by a process different from that for the surface
outside the border, so there is no reason to expect that the two
slopes join smoothly.  It is thus natural to expect that a sudden
change in the slope is observed at the border between the solid-like
and fluid-like regions.

In summary, we have discovered circular kinks on the surface of
granular material in a spinning bucket when its axis of rotation is
tilted beyond the angle of internal friction. The radius of the kinks
depends on the angular velocity, the tilting angle and the angle of
internal friction. With a fixed tilting angle, the dimensionless
radius of kinks $R_k = r_k \omega^2 /g$ remains roughly constant. We
find that the surface is divided into two regions: a fluid-like inner
and a solid-like outer regions. We determine the critical radius $R_c$
from the radial stability condition and the prediction reflects the
basic features of the experiments. Using a simple cellular automata
model, with the same stability condition and by allowing the
propagation effect of avalanche, we obtain the granular surface in
good accord with the experiments.

This work is supported by the Basic Science Research Institute, Seoul
National University and by Korea Science and Engineering Foundation
(KOSEF) through the Science Research Center for Dielectrics and
Advanced Matter Physics. One of us (J.L.) is supported in part by
KOSEF through the Brain-pool program and SNU-CTP.\\

$^{\dagger}$Electronic address : isyu@snu.ac.kr

\begin{figure}
\centerline{\psfig{figure=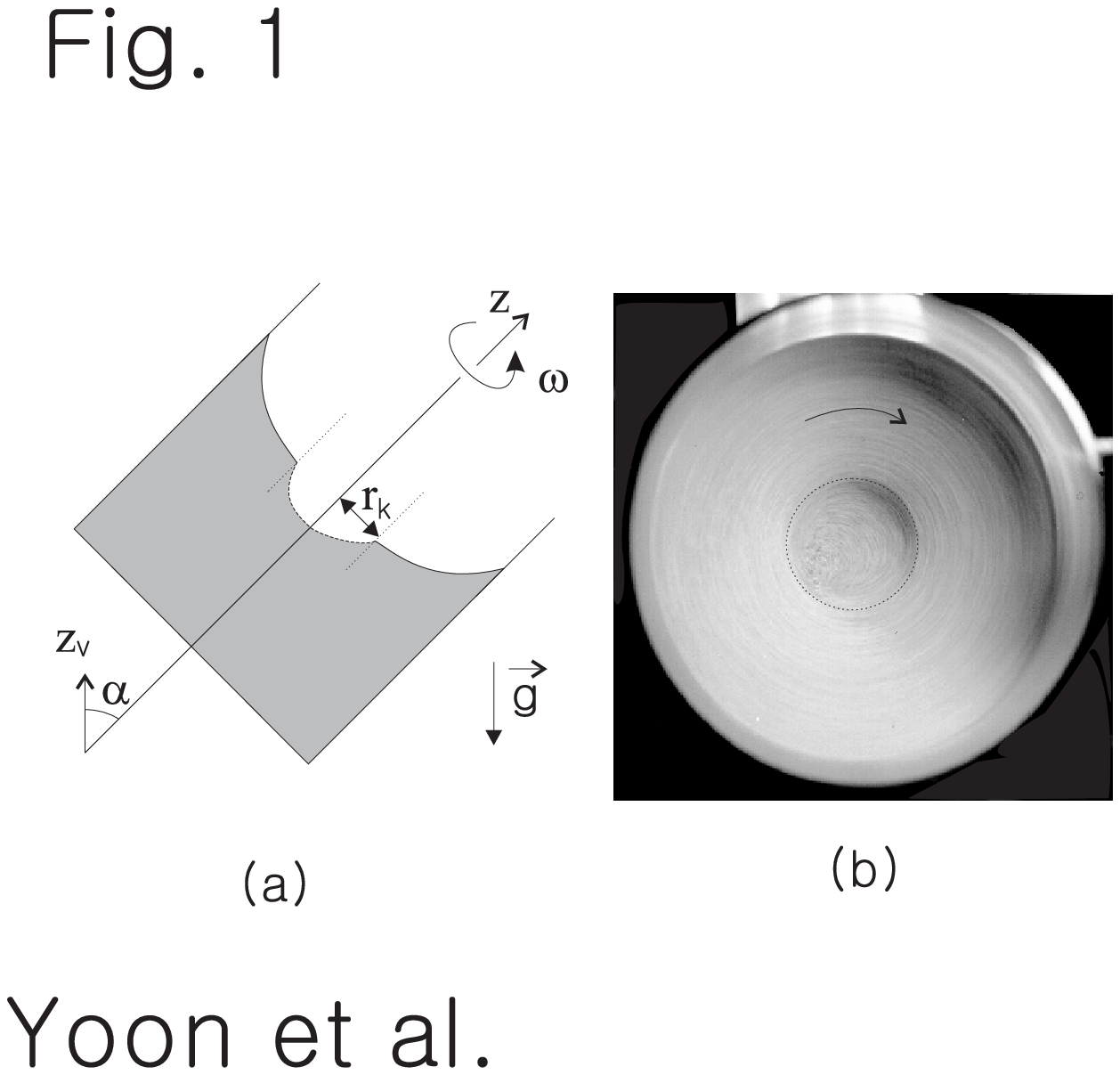,angle=0,width=6in}}
\vfill

\caption{(a) Schematic side-view of a tilted spinning bucket: the
rotational axis $z$ is tilted from the vertical direction
$z_{\mbox{v}}$ by an angle $\alpha$. A kink of radius $r_k$ forms on
the granular surface separating the solid-like (profile stable) outer
region from the fluid-like (profile unstable) inner region. $\vec{g}$
denotes the gravity. (b) Top-view photograph of a kink formed on the
sand surface (stressed by a dotted circle, of radius 2.0 cm, for a
clearer view) rotated with $\omega$ = 230 rpm (direction is indicated
by an arrow) and $\alpha$ = 45$^{\circ}$.}
\end{figure}
\newpage

\begin{figure}
\centerline{\psfig{figure=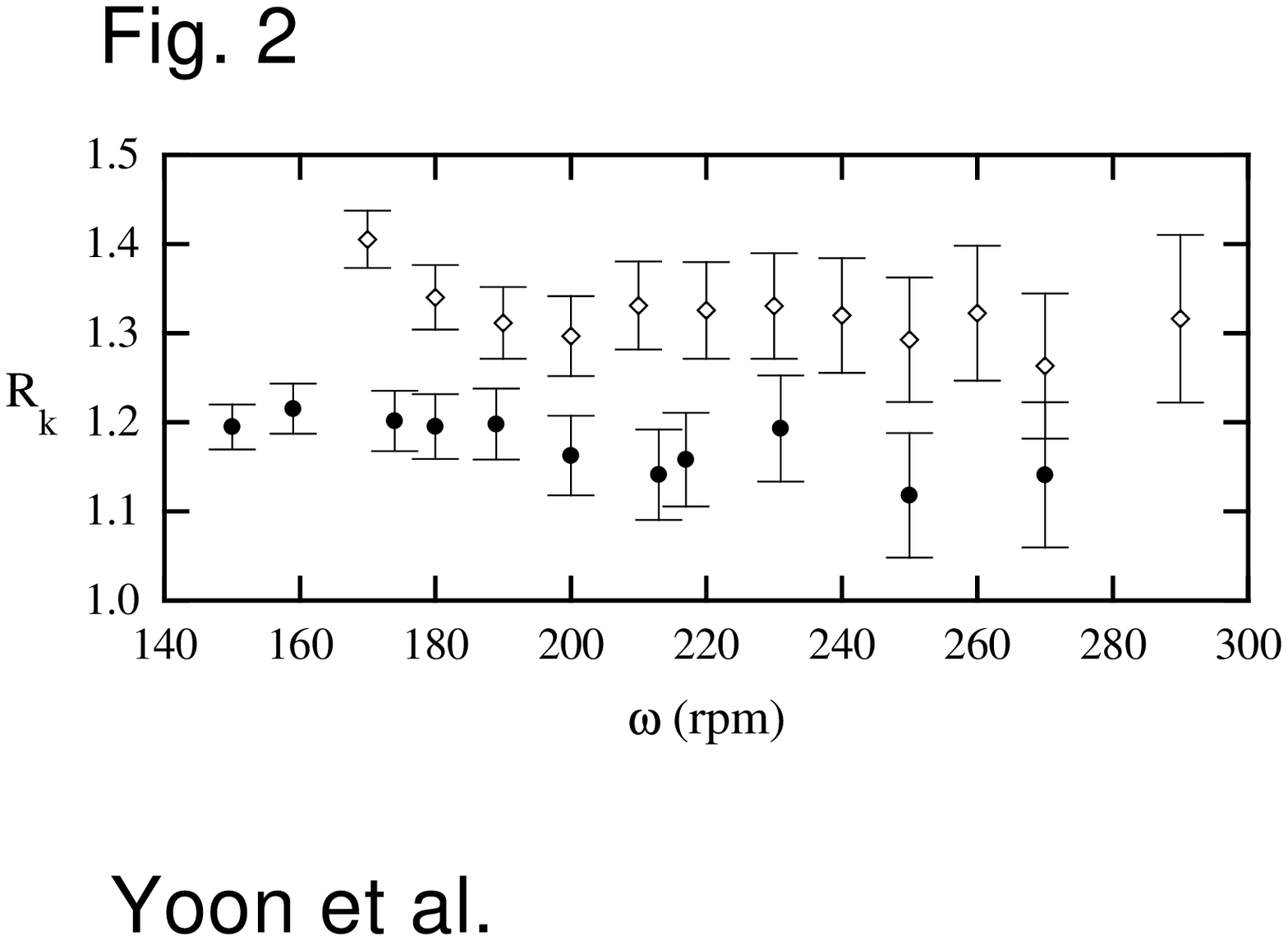,angle=0,width=6in}}
\vfill

\caption{The dimensionless radius $R_k=\frac{r_k \omega^2}{g}$ of the
kinks formed on the sand pile. Examples for the two tilting angles,
$\alpha=60^{\circ}$ ($\diamond$) and $\alpha=45^{\circ}$ ($\bullet$),
indicate the relation $r_k \propto \omega^{-2}$.  Errors of the
measurements are $\Delta r_k$ $\simeq$ 1 mm.}
\end{figure}
\newpage

\begin{figure}
\centerline{\psfig{figure=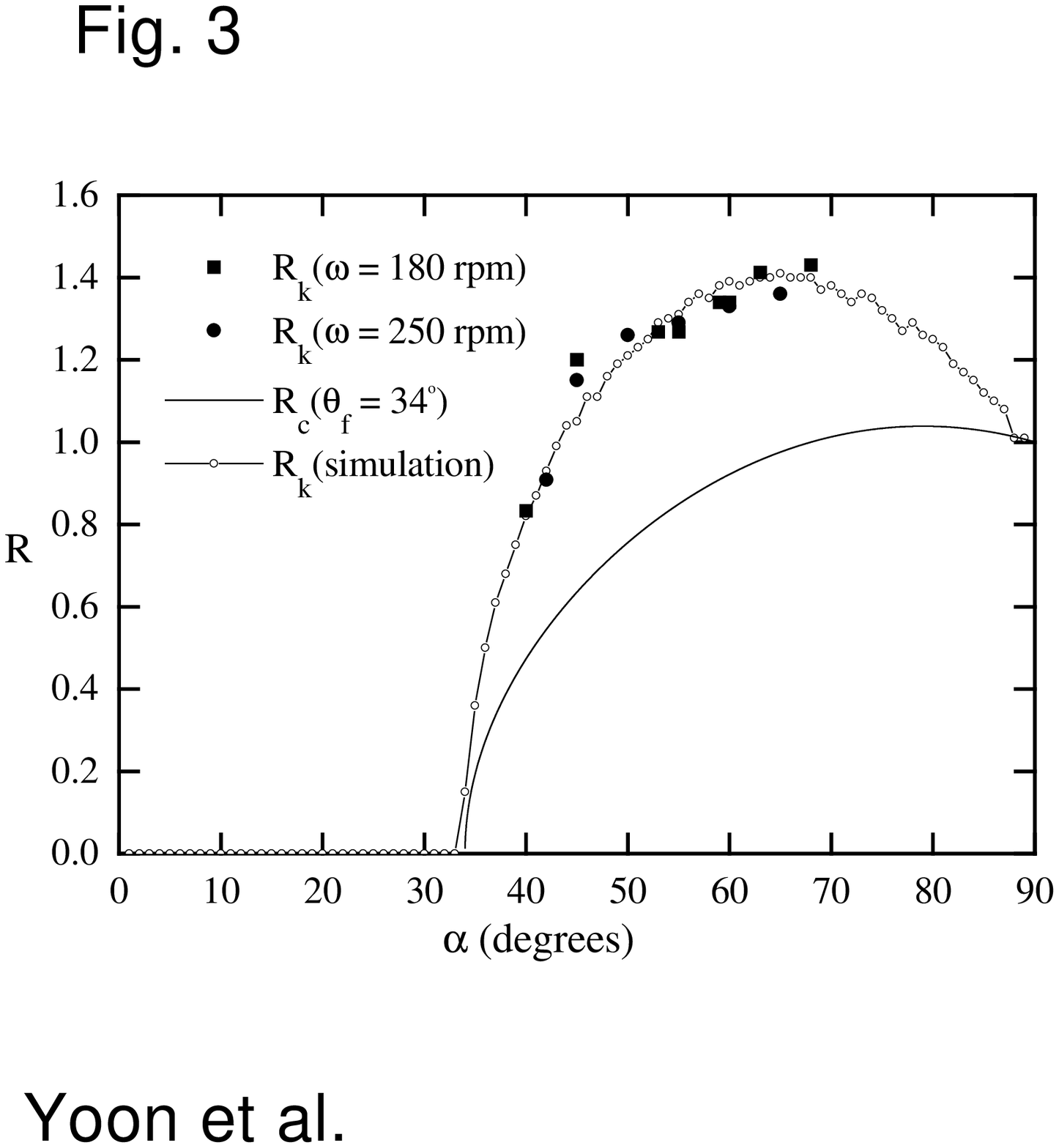,angle=0,width=5in}}
\vfill

\caption{The dimensionless radius $R_k $ for sand at various $\alpha$.
Representative data for two values of $\omega$ are shown together with
the critical radius $R_c$(------) calculated for stability and the
$R_k $ from simulation($-\circ-$) with the angle of internal friction
(angle of repose) $\theta_f =34^{\circ}$.}
\end{figure}
\newpage

\begin{figure}
\centerline{\psfig{figure=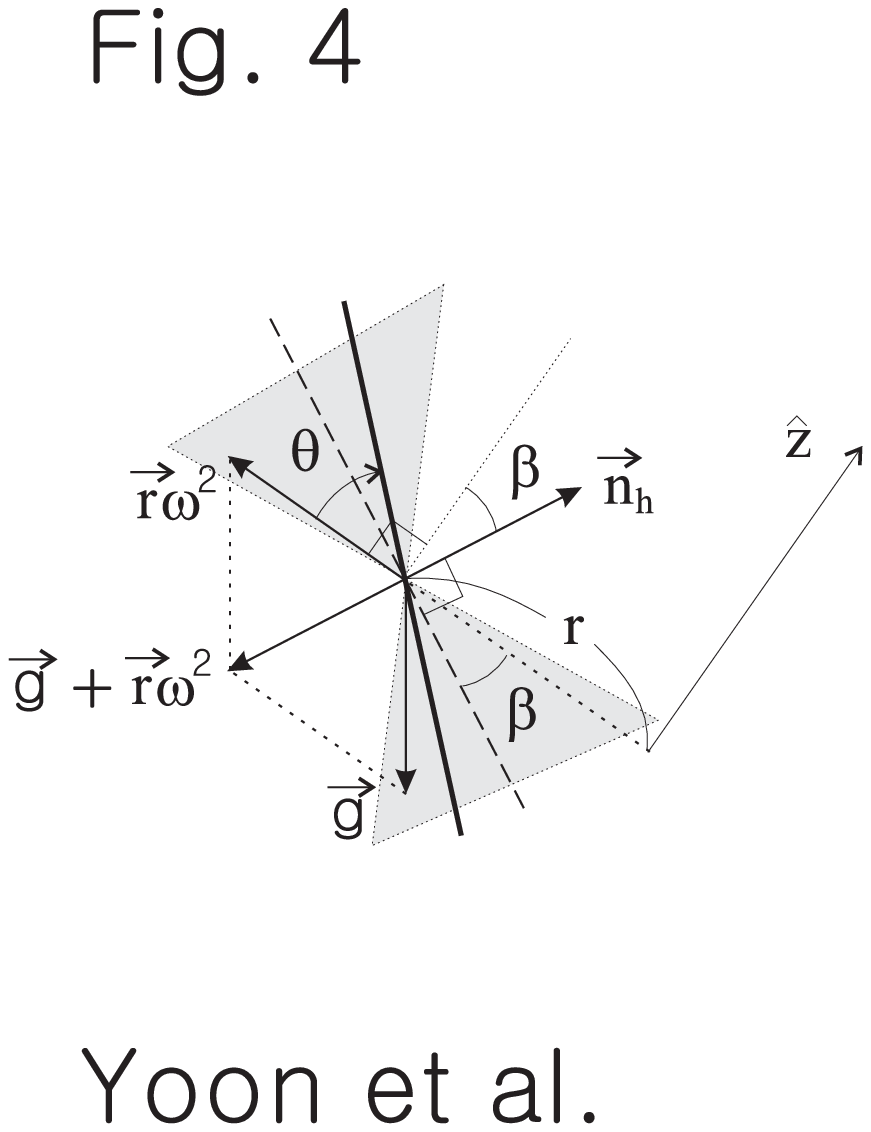,angle=0,width=6in}}
\vfill

\caption{The vector $\vec{n}_h $ is normal to the local surface
(dashed line) when there is no friction. The angle $\beta$ between the
vector $\vec{n}_h $ and the rotation axis $\hat{z}$ is changing during
rotation. The granular surface (solid line) in the shaded region is
stable for the angle $\theta$ between the granular surface and bottom
of the bucket in the range $\beta -\theta_f \leq \theta \leq \beta +
\theta_f $, where $\theta_f $ is the angle of repose. This figure is
an example that a grain is at the top position during rotation.}
\end{figure}
\newpage

\begin{figure}
\centerline{\psfig{figure=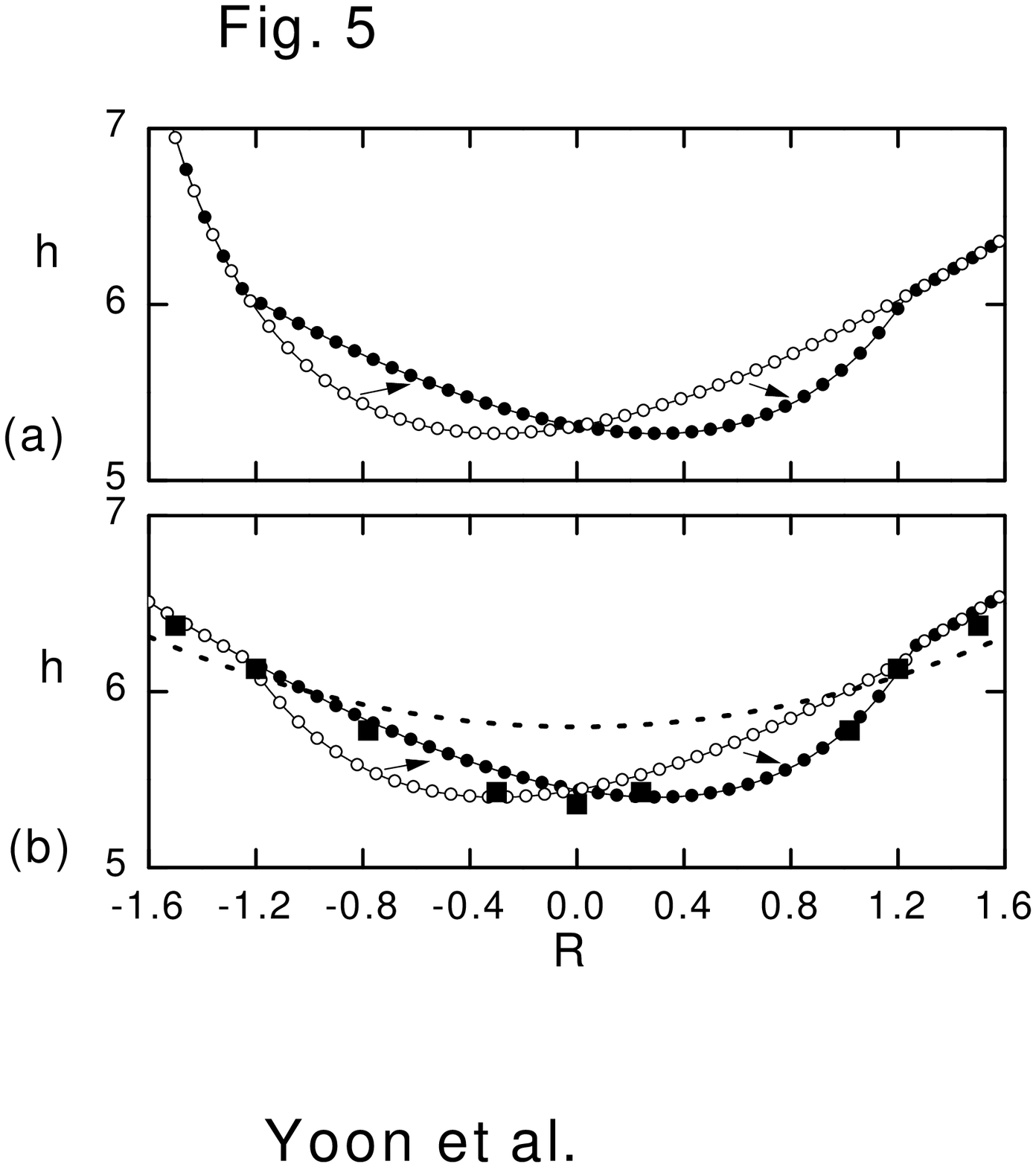,angle=0,width=5in}}
\vfill

\caption{Time evolutions of the granular surface calculated from the
simulation of a simple cellular automata model with
$\theta_f=34^{\circ}$ and $\alpha = 45^{\circ}$ for two different
initial conditions; (a) first tilt and then rotate, (b) first rotate
vertically and then tilt slowly. Different initial preparations give
different surface profiles, which however result in the same radii of
kinks. The open and filled circles indicate the surface profiles at
the initial and after a half ($\pi$-radian) rotation positions,
respectively, and two arrows represent the direction of changes. In
(b), the filled squares are the experimental data, which are scaled by
$\frac{\omega^2}{g}$, and the dashed line is the parabolic best fit to
the data in the solid-like region ($1.2 < |R| <4$). The experimental
data in the fluid-like region ($|R| < 1.2$) are only rough estimates
($\Delta h$ $\simeq$ 0.2) due to the steady flow of grains. The data
at $R > 0$ correspond to the upper half of the granular surface.}
\end{figure}
\newpage

\end{document}